\documentclass[prb,twocolumn,amsmath,amssymb,floatfix,aps,showkeys,showpacs]{revtex4}
\usepackage{graphicx}
\usepackage{dcolumn}
\usepackage{bm}

\usepackage[normalem]{ulem}

\begin{document}

\title{Critical behavior of the Random-Field Ising Magnet
 with long range correlated disorder}

\author{Bj\"orn Ahrens}
\email{bjoern.ahrens@uni-oldenburg.de}
\author{Alexander K. Hartmann}
\affiliation{Institute of Physics, University of Oldenburg, 26111 Oldenburg, 
Germany}

\date{\today}

\begin{abstract}
We study the correlated-disorder 
driven zero-temperature  phase  transition of the Random-Field
Ising Magnet using exact numerical 
ground-state calculations for  cubic lattices.
We consider correlations of the quenched disorder 
decaying proportional to $r^{a}$, where $r$ is the
distance between two lattice sites and $a<0$.
To obtain exact ground states, we use a well established mapping to
the graph-theoretical maximum-flow problem, which
allows us to study large system sizes of more than two million spins.
We use finite-size scaling analyses for values    
$a=\{-1,-2,-3,-7\}$ to calculate the critical point 
and the critical exponents characterizing the
behavior of the specific heat,  magnetization, 
susceptibility and of the correlation length close to the 
critical point.  We find basically the same critical behavior as for the RFIM 
with $\delta$-correlated disorder, except for the finite-size exponent of the susceptibility and for the
case $a=-1$, where the results are also compatible with
a phase transition at infinitesimal disorder strength.
 A summary of this work can be found at the {\tt papercore} database {\tt www.papercore.org}.
\end{abstract}


\maketitle
\section{\label{Introduction}Introduction}
The random-field Ising Magnet (RFIM) is a prototypical model 
for magnetic systems with quenched disorder. For $d = 3$ and higher 
dimensions, \cite{bricmont1987} it is known to undergo a second-order phase transition  \cite{GofmanAdlerAharonyHarrisSchwartz1993,Rieger1995,Nowak1998,art_uli1999, HartmannYoung2001,middleton2002,frontera2002,seppala2002,Hartmann2002, middleton2002b,fes_rfim2008,AhrensHartmann2011} at a critical temperature $T_c$ or disorder strength $h_c$ 
($T_c(h)\Leftrightarrow h_c(T)$): 
For low temperatures and weak disorder  the ferromagnetic interactions dominate and the system is long-range ordered. 
For large  temperature or  strong disorder, the RFIM exhibits no long-range order and behaves like a paramagnet in a field.

The quenched disorder used in earlier studies of the RFIM was 
mostly uncorrelated ($\delta$-correlated).\cite{GofmanAdlerAharonyHarrisSchwartz1993,Rieger1995,Nowak1998,art_uli1999, HartmannYoung2001,middleton2002,frontera2002,seppala2002,Hartmann2002, middleton2002b,fes_rfim2008}
This is quite common in the literature when
studying disordered systems like percolation, random ferromagnets,
spin glasses or polymers in random media. Nevertheless, real systems
are always emerging from physical processes, hence correlations are present,
which could play an important role for its behavior.
Here, we consider a tunable, scale-free (power-law), i.e. long range, correlation to the random field to explore its influence on the 
critical behavior. Please note that for an exponentially decreasing
correlation strength with a typical length scale $\Xi$, 
via renormalizing the system beyond $\Xi$, the behavior of the uncorrelated system should
be recovered.
The $O(n)$ random-field model with long-range correlated disorder 
was studied recently  \cite{fedorenko2007} via functional renormalization
group methods around $d= 4$ and for values $n>3$, i.e., 
without including the Ising case $n=1$.
For other types of random systems, there exist already some studies
for the case of long-range correlated disorder, e.g., for percolation,\cite{MakseHavlin1996}
the diluted Ising ferromagnet,\cite{BallesterosParisi1999}
random walks,\cite{hod2004} or elastic systems.\cite{fedorenko2008}

Now, we state our model in detail.
The RFIM consists of Ising $N=L^3$ spins $S_i=\pm 1$ 
located on the sites of a cubic lattice with periodic boundary conditions
in all directions. The spins couple to each other and to local net fields. Its Hamiltonian reads 
\begin{equation}
\mathcal{H}=-J\sum_{\langle i,j\rangle}S_iS_j - \sum_i\left(h \eta_i +H\right)S_i\,.
\end{equation}
It has two contributions. The first covers the spin-spin interaction, 
where $J$ is the ferromagnetic coupling constant between two 
adjacent spins and $\langle i,j\rangle$ denotes pairs of
next-neighboured spins.
The second part of the Hamiltonian describes the coupling to local, 
and global fields $h \eta_i$ and $H$, respectively. 
The factor $h$ is the disorder 
strength used to trigger the phase transition. 
The global field is included only for technical reasons
to calculate the susceptibility in the limit $H\to 0$.
 The quenched local fields $\eta_i$ are Gaussian distributed with 
zero mean and unity width. The important property of these fields 
is their spatial long-range correlation. It decays as a power law 
\begin{equation}
\mathcal{C}(\vec r) \equiv
\Big\langle \frac 1 N \sum_{\vec x}\eta(\vec x) \eta(\vec x + \vec r)
\Big\rangle \sim  |\vec r|^a
\label{eq:def:C}
\end{equation}
 with a tunable, well defined decay exponent $a$. 
The symbol $\langle \;\ldots\; \rangle$ denotes 
the average over the quenched disorder and  $\vec x$ are the 
positions of a lattice sites $i$. 

We will study in particular the values $a=\{-1,-2,-3,-7\}$.
First of all, it is interesting to know whether this type of disorder
is relevant with respect to the ordered case.
A hint to the answer of this question comes from
the case of systems with  ``random-temperature disorder'',
like the diluted ferromagnet:
For a $d$-dimensional system, if $|a|>d$, i.e. when
the disorder correlation vanishes rather quickly, the
usually Harris criterion applies.\cite{WeinribHalperin1983} 
The Harris criterion \cite{harris1974}
states that the disorder is relevant if $d\nu-2<0$, $\nu$ being
the critical exponent of the ordered system. For the $d=3$
ferromagnet, we have $\nu=0.6294(5)$ from 
Ref.\ \onlinecite{ballesteros1999}, hence
the disorder is relevant, as known from the case of uncorrelated
disorder. For $|a|<d$, in particular for $a=-1$ and $a=-2$ as studied
here, the disorder is relevant according to 
Ref.\ \onlinecite{WeinribHalperin1983}
for $2/|a|>\nu$. Since $2/1=2>0.6294$ and $2/2=1>0.694$, the disorder
will be relevant also for these values of $a$. 
The results we present below in this work, although for a different
type of disorder, are compatible with these predictions.
 Furthermore, we will consider
the question whether the correlated disorder is different from the
behavior of the uncorrelated disorder case. Our results show that
the most exponents are compatible within error bars
with the values of the standard RFIM, but the combination
$\gamma/\nu$ shows a clear signature of non-universality. 
This is similar to the diluted Ising model,
where the long-range correlated dilution clearly changes some but not all 
critical exponents \cite{BallesterosParisi1999} 
with respect to the uncorrelated case. \cite{ballesteros1998}

The paper is organized as follows: In section \ref{num_meth} we sketch
 the idea how to calculate Gaussian distributed correlated random numbers. 
After that a brief description of  the numerical ground-state
approach is given. The measured quantities and the methods to 
analyze the data are displayed in section \ref{QI}.  
Our numerical results are presented in section \ref{Results}. Based on the results we discuss the extremes of correlated disorder. The last section contains the discussion and conclusion. An extensive summary of this work ($1/10$ of the length) can be downloaded
from the \emph{Papercore} database.\cite{papercore_gen}
\section{Numerical methods\label{num_meth}}

In this section, we first explain how we generated the samples of the
correlated disorder. Second, we briefly outline the numerical approach to calculate the exact ground
states of these samples.

To obtain a realization of correlated random fields, we basically
apply the ideas of Refs.\
\onlinecite{WeinribHalperin1983,MakseHavlin1996,BallesterosParisi1999}.
The recipe is, to demand for a convolution kernel $\Phi(\vec r)$ which
convolves iid random numbers $u(\vec r)$, such that $\eta(\vec r)=\Phi(\vec r) \ast u(\vec r) = \sum_{\vec x} 
\Phi(\vec x)  u(\vec r-\vec x)$ (using periodic boundary conditions
for $u(.)$) show a desired two point correlation. Power law correlations are created, using 
\begin{align}
 \mathcal{C}(\vec r) = \ (1 + |\vec r|^2)^{a/2}\hspace{1em} a < 0 \label{CorrFun}.
\end{align}
The long range behaviour is the same as of a pure power law without a singularity at the origin. This avoids zero-mode
divergence.\cite{Peng1991,Prakash1992}

 In Fourier space, the transformation $\mathcal{F}$ given through
 $\tilde\eta(\vec k)\equiv \sum_{\vec x}e^{i \vec k \cdot \vec x} \eta(\vec x)$,
the correlation function is equivalent to the spectral density. Applying the definition of $C(\vec r)$ from Eq.\ (\ref{eq:def:C}) results to
\begin{align}
\langle \tilde\eta(\vec k) \tilde\eta^*\!(\vec k) \rangle
= & \Big \langle \sum_{\vec r} e^{i \vec k \cdot (\vec x + \vec r)}
 \eta(\vec x + \vec r)
\sum _{\vec x} e^{-i \vec k \cdot \vec x} \eta (\vec x) \Big\rangle
 \nonumber \\
= & N \tilde{ \mathcal{C}}(\vec k)\,.
\label{PowSpec}
\end{align}
A convolution in real space turns to a multiplication in Fourier
space:
\begin{align}
 	\tilde\eta(\vec k)&= \tilde\Phi(\vec k) \tilde u(\vec k)
 	\label{FTeta}.
\end{align}
Now insert Eq.(\ref{FTeta}) into Eq.(\ref{PowSpec}) to
determine the convolution kernel.
\begin{align}
N \tilde{ \mathcal{C}}(\vec k) = \langle \tilde\Phi(\vec k) \tilde
u(\vec k) \tilde\Phi^*\!(\vec k) \tilde u^*\!(\vec k) \rangle  =
|\tilde\Phi(\vec k)|^2 \langle |\tilde u(\vec k)|^2 \rangle.
\end{align}
We choose the real space random numbers $u(\vec x)$ as being
distributed iid according a Gaussian with zero mean and variance one,
\begin{eqnarray}
\langle u(\vec x ) \rangle & = & 0 \\
\langle u(\vec x) u (\vec y) \rangle & = & \delta_{\vec x, \vec y}\,,\
\end{eqnarray}
 such that the variance $\langle |\tilde u(\vec k)|^2 \rangle = N$.
This results in $\tilde{ \mathcal{C}}(\vec k) =|\tilde\Phi(\vec k)|^2$,
such that we can choose the correlated random numbers (in the Fourier space).
\begin{align}
\tilde\eta(\vec k)&= \sqrt{\tilde{ \mathcal{C}}(\vec k)} \tilde u(\vec k)
\end{align}
The back transformed  correlated random numbers $\eta(\vec r)$ are real numbers. Since $u(\vec r) \in \mathbb{R}$,  $\tilde u(-\vec k) = \tilde u^*\!(\vec k)$. From Eq.\ (\ref{PowSpec}) we infer $\tilde{\mathcal{C} }(\vec k)=\tilde{\mathcal{C}} ( | \vec k | ) \in \mathbb{R}^+$. So the back transformation $\eta(\vec
r)=\mathcal{F}^{-1}(\tilde\eta(\vec k))(\vec r) \in \mathbb{R}$.

\begin{figure}[th]
\begin{center}
\includegraphics[width=0.5\textwidth]{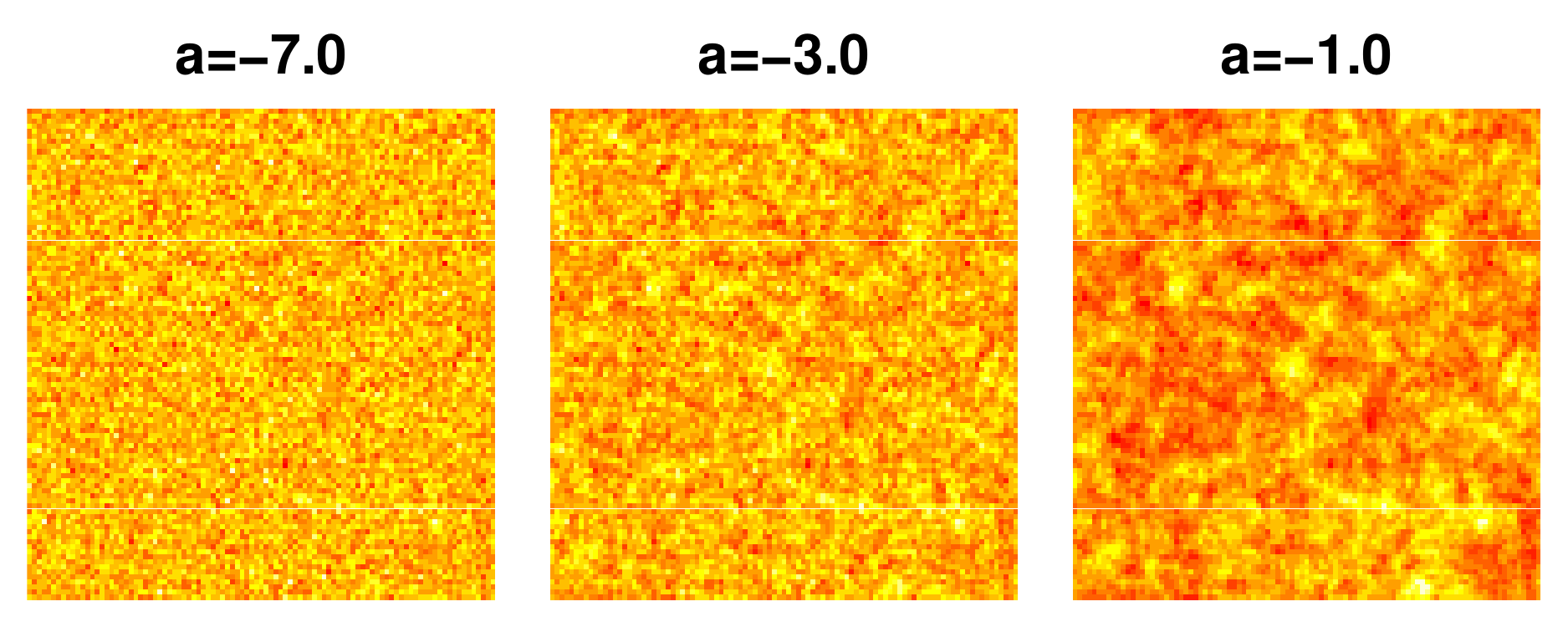}
\caption{(color online) Slices of the same correlated disorder of a
$97^3$ lattice for different correlation strengths. The random fields
are heat coded. Bright means high positive field and dark means high
negative field. \label{3corrExample}}
\end{center}
\end{figure}

For maximum flexibility, to test different types of correlation
functions, we have implemented the Fourier transformation numerically,
using the  \emph{Fastest Fourier Transform in the West} (FFTW) library, version 3.2.2.\ \cite{FFTW05} An
example of a realisation of the correlated disorder is shown in Fig.\ \ref{3corrExample}
 for different values of $a$.

\begin{figure}[htp]
\begin{center}
\includegraphics[width=0.5\textwidth]{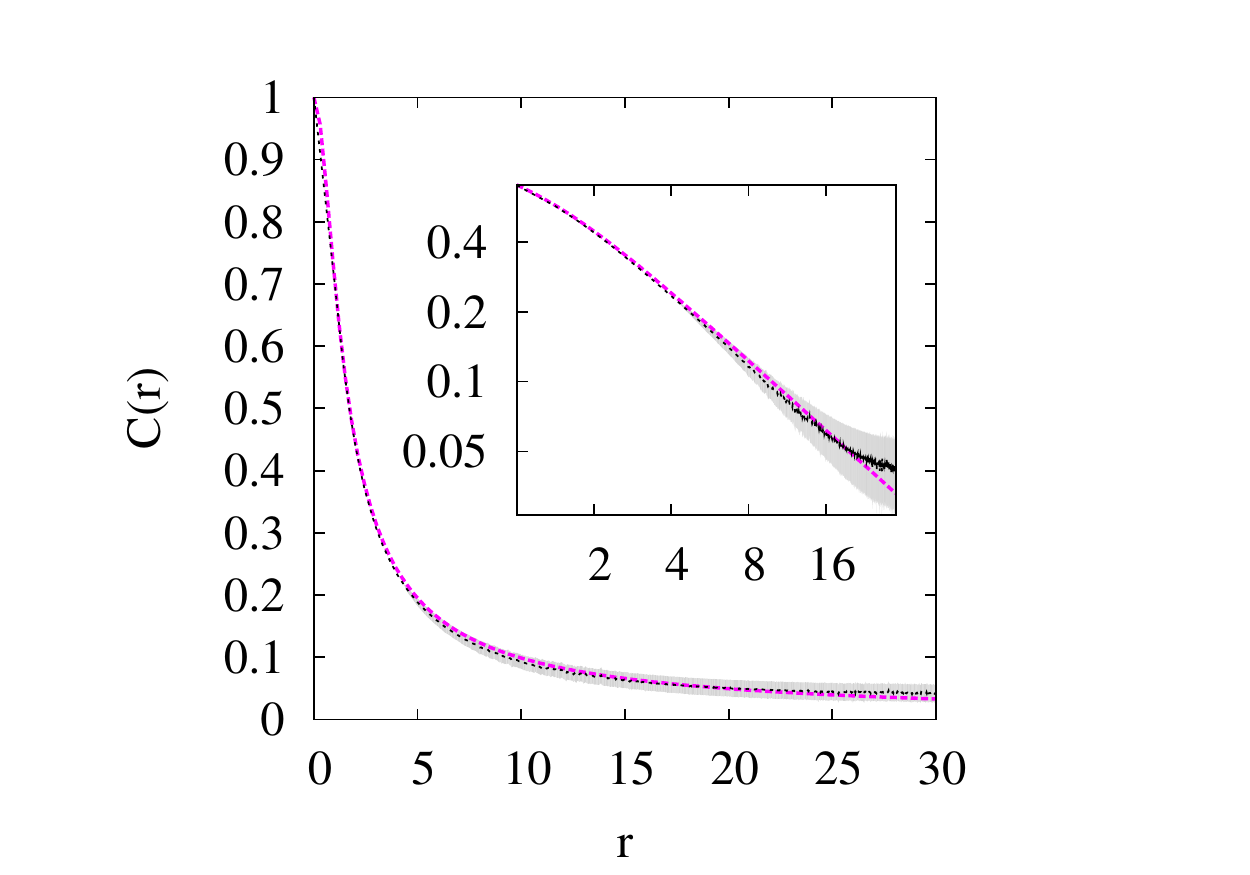}
\caption{(color online) Mean of the two-point correlations for 20
arbitrary samples of long-range correlated random fields for
$L=49,a=-1$ (black line) and error (gray
background). The dashed (magenta) line is a fit according to Eq.\ (\ref{CorrFun}) with $a=1.003(2)$ . The inset shows the same data as log-log-plot.
\label{correlation_L49_G-1}}
\end{center}
\end{figure}

We tested our procedure by generating 20 realizations for
$a=-1$ of the correlated disorder,  calculating the two-point correlation
Eq.\ (\ref{eq:def:C}) directly, and fitting Eq.\ (\ref{CorrFun}) with variable
exponent $a$.  The correlation, shown in Fig.\ \ref{correlation_L49_G-1},
and the resulting value $a=1.003(2)$ 
show that except for very small correlation
at large distance the procedure works very well.

\begin{figure}[th]
\begin{center}
\includegraphics[width=0.25\textwidth, angle=270]{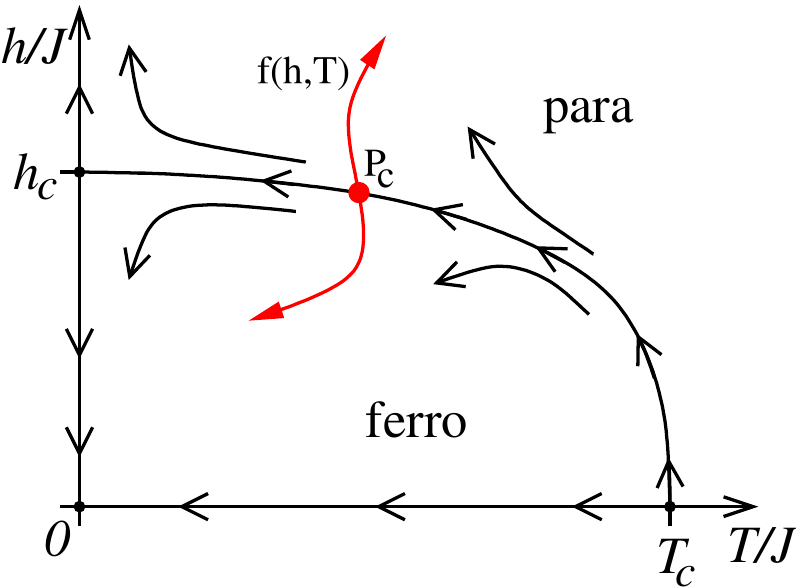}
\caption{(color online) Schematic diagram of the phase space of the RFIM. The path $f(h,T)$ (red line) shows an arbitrary path to cross the phase boundary. The small arrows denote the renormalization group flow.\label{PhaseDiagramm}}
\end{center}
\end{figure}

Next, we mention shortly how the exact ground states are calculated.
The phase space of the uncorrelated RFIM consists of a ferromagnetic and a
paramagnetic phase, see Fig.\ \ref{PhaseDiagramm}. 
The transition from one phase to the other can be triggered by varying the disorder
strength $h$ or the temperature $T$.  Changing both along a path
$f(h,T)$ in the phase space leads to a critical point
$P_c=(h_c,T_c)_{f(h,T)}$. From renormalization group calculations, the
RFIM is known \cite{BrayMoore1985} 
to exhibit the same critical behavior at any $P_c$,
 except the temperature-driven phase transition point
of the standard non-random Ising model. Hence, it is possible
to focus on  $T=0=\text{const.}$ and vary
just $h$, to study the critical behavior along the full
transition line. We do not know a-priori, whether the phase
dia\-gram for the correlated-disorder case has the same property,
nevertheless, it makes sense to concentrate, at least for our
study presented here, also on $T=0$. 
From the computational point of view this is very favorable, since it
is possible to calculate exact ground states at $T=0$ in a very
efficient way for system sizes as large as  $N=141^3$ spins.
Within this approach \cite{PicardRatliff1975,ogielski1986},
each realization of the correlated net fields $( \{\eta_i\},H)$ has to
be mapped to a graph with $N+2$ nodes and $2N+1$ edges with suitable
edge capacities. On this graph a sophisticated maximum flow/minimum
cut algorithm can be applied.\cite{GoldbergTarjan1988,HartmannRieger2001} The
resulting minimum cut directly correspond to the GS spin configuration
$\{S_i\}$ of that specific realization of the net disorder. We
used the efficient maximum-flow subroutines implemented
in the LEDA library.\cite{leda1999}

\section{\label{QI}Quantities of interest}
 From a GS spin configuration, some quantities of interest can be obtained directly, as the magnetization per spin 
\begin{equation}
 M=\frac{1}{N}\sum_i^N S_i\;
\end{equation}
and the bond energy per spin
\begin{equation}
E_J=\frac{J}{N}\sum_{\langle i, j \rangle} S_i S_j\;,
\end{equation}
Using these individual values, we calculate averaged quantities like 
the average magnetization $m=\langle M\rangle  $. This 
disorder average $\langle \ldots \rangle$ is performed always for a fixed
 value of $h$. We also consider the Binder cumulant\ \cite{Binder1981}
\begin{equation}
g(h,L) = \frac{1}{2}  \left(  3 - \frac{  \left\langle  M^4  \right\rangle }
{  \left\langle M^2  \right\rangle ^2_h}   \right).
\label{eq:Binder}
\end{equation}
A specific-heat-like quantity $C(h)$ can be calculated as the
numerical derivative of $E_J$ with respect to $h$ (see Ref.\
\onlinecite{HartmannYoung2001} for details). From here on we will
refer to it as specific heat.
\begin{equation}
 C(h)= \frac{\left\langle \partial E_J(h)\right\rangle}{\partial h}\;.\label{SpecificHeat}
\end{equation}
We also calculated the zero-temperature susceptibility
\begin{equation}
\chi(h)=\left. \frac{\partial m(h,H)}{\partial H}\right|_{H=0}\;\label{SuszDef}
\end{equation}
 as linear response of the magnetization to small homogeneous magnetic fields $H$. Therefore, we apply small homogeneous fields at equidistant values $H_1,2H_1,3H_1$ and fit parabolas as function of $H$ to the  magnetizations $m(h,H=0)$,  $m(h,H_1)$,  $m(h,2H_1)$ and  $m(h,3H_1)$. For a fixed value of the disorder strength $h$ the linear coefficient corresponds to the  susceptibility $\chi(h)$.
 
We will see that the results are compatible with second order
phase transitions, such that the measured quantities show power-law
behavior close to the phase transition point. To determine the critical
exponents, we use the standard scaling forms, i.e.,
 \begin{alignat}{2}
g(h,L) & = &
\tilde g\left((h-h_c)L^{1/\nu}\right)\,,\label{binderRescale} \\
m(h,L) 	&= L^{-\beta/\nu} 
	&\tilde m\left((h-h_c)L^{1/\nu} \right)\,,\label{magRescale} \\
\chi(h,L) &= 
L^{\gamma/\nu} &\tilde \chi\left( (h-h_c)L^{1/\nu}\right)\,,\label{chiRescale} \\
C(h,L) &=  
L^{\alpha/\nu} &\tilde C\left ( (h-h_c)L^{1/\nu} \right)\,, \label{specHRescale} 
\end{alignat}
and apply a finite-size scaling analysis. For the Binder cumulant and
the magnetization we use a nice tool which performs data collapses
automatically.\cite{autoScale2009} It is based on a simplex
algorithm and is written in python.

The specific heat and the
susceptibility show a maximum close to the critical point. 
at some argument $f$ of the universal functions $\tilde \chi(\cdot)$ and $\tilde C(\cdot)$. 
Note that the peak positions for specific heat and susceptibility 
of the same system sizes $L$ usually differ. Thus, also the value of $f$
(and even the sign) may differ.  From  Eqs.\ (\ref{chiRescale}) and  (\ref{specHRescale})
it follows that the finite-size dependence of the positions of the maxima,
respectively, scale as
\begin{equation}
h^*(L) = h_c + f\cdot L^{-1/\nu} \label{finite_hin_f}.
\end{equation}
Furthermore,  right at $h^*(L)$, the height of the maxima should scale as $L^{\gamma/\nu}$ and $L^{\alpha/\nu}$. In the
case of $\alpha=0$ other forms like a logarithmic divergence or a 
convergence to a constant (``cusp'') have been
observed for other systems in the 
literature.\cite{yeomans1993,HartmannYoung2001} 
Below we present the results we obtained for the position and the height 
of the peaks and test their scaling behaviour according to these scaling 
assumptions.

As mentioned above, these quantities are average values. They are
strongly dependent on the set of disorder realizations taken into
account. Hence, we perform an average usually over many thousands of realizations.
We estimate the variability of these average values from $200$ bootstrap samples\ \cite{bootstrapbook,Hartmann2009} 
and quote this as error.  
\section{\label{Results}Results}

We performed exact ground-state calculations
for three-dimensional RFIMs for correlation strengths 
$ a=\{-1,-2,-3,-7\} $. We considered system sizes ranging 
from $L=7$ to $L=141$.
The number of disorder realizations per system size and
correlation strength can be found in Tab.\ \ref{samples}. The actual
number of calculated ground states is four times larger, since $4$
different external fields are needed to obtain a susceptibility. The values of $H_1$ are stated in the very right column of Tab.\ \ref{samples}.

\begin{table}
 \begin{tabular}{| r| r| r| r| r| c|}\hline
 $L$		&$a=-1$ &$a=-2$	&	$a=-3$	&	$a=-7$	&$H_1$\\\hline
$7$  	&$33$	&$33$	&$	33$		&$	33 $		&$ 0.0075$\\\hline
$11$	&$33$	&$33$	&$	33$		&$	33$		&$ 0.0050$\\\hline
$15$	&$13$	&$32$	&$	61$		&$	182$	&$ 0.0030$\\\hline
$21$	&		&$29$	&$	14$		&$	155$	&$ 0.0025$\\\hline
$25$	&$90$	&$32$	&			&$	150$	&$ 0.0015$\\\hline
 $35$	&$44$	&$10$	&$	45$		&$	14$		&$ 0.0015$\\\hline
$49$	&$32$	&$26$	&$	16$		&$	16$		&$ 0.0010$\\\hline
$69  $	&$16$	&$24$	&$	9$		&$	15$		&$ 0.0005$\\\hline
$97$	&$20$	&$8$	&$	  4.5$	&$	9.5$		&$ 0.0004$\\\hline
$117$	&$3$	&$1$	&$	1.2$		&$	1.8$		&$ 0.0002$\\\hline
$141$	&$1$	&$0.2$	&$	0.2$		&$	0.3$		&$ 0.0001$\\\hline
 \end{tabular}
\caption{Number of disorder realizations per system size and correlation strength in thousands ($10^3$) and the external field $H_1$, used to calculate the susceptibility. \label{samples}}
\end{table}

 \begin{figure}[th]
\begin{center}
\includegraphics[width=0.5\textwidth]{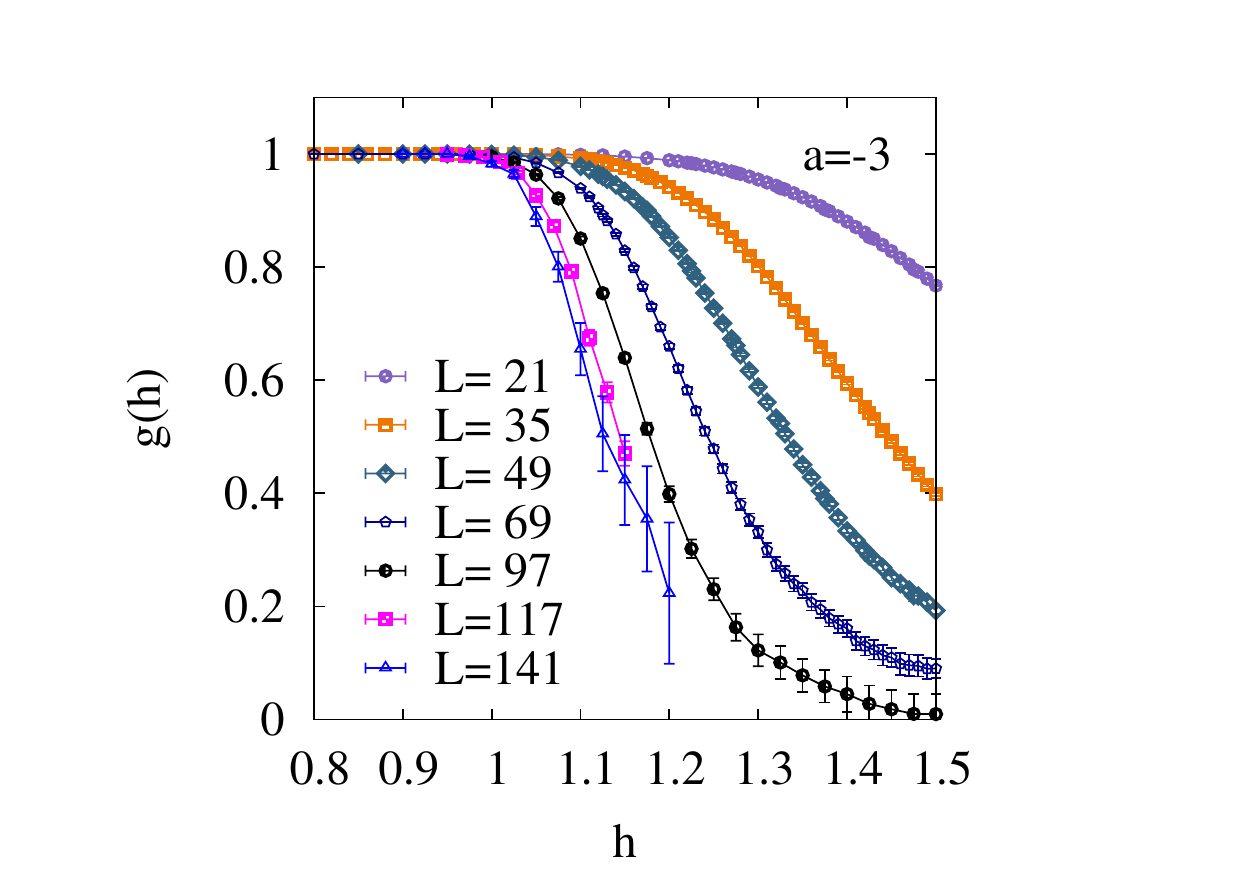}
\caption{(color online) 
The Binder cumulant for $a=-3$ and different system sizes.
\label{fig:binder}}
\end{center}
\end{figure}
 \begin{figure}[th]
\begin{center}
\includegraphics[width=0.5\textwidth]{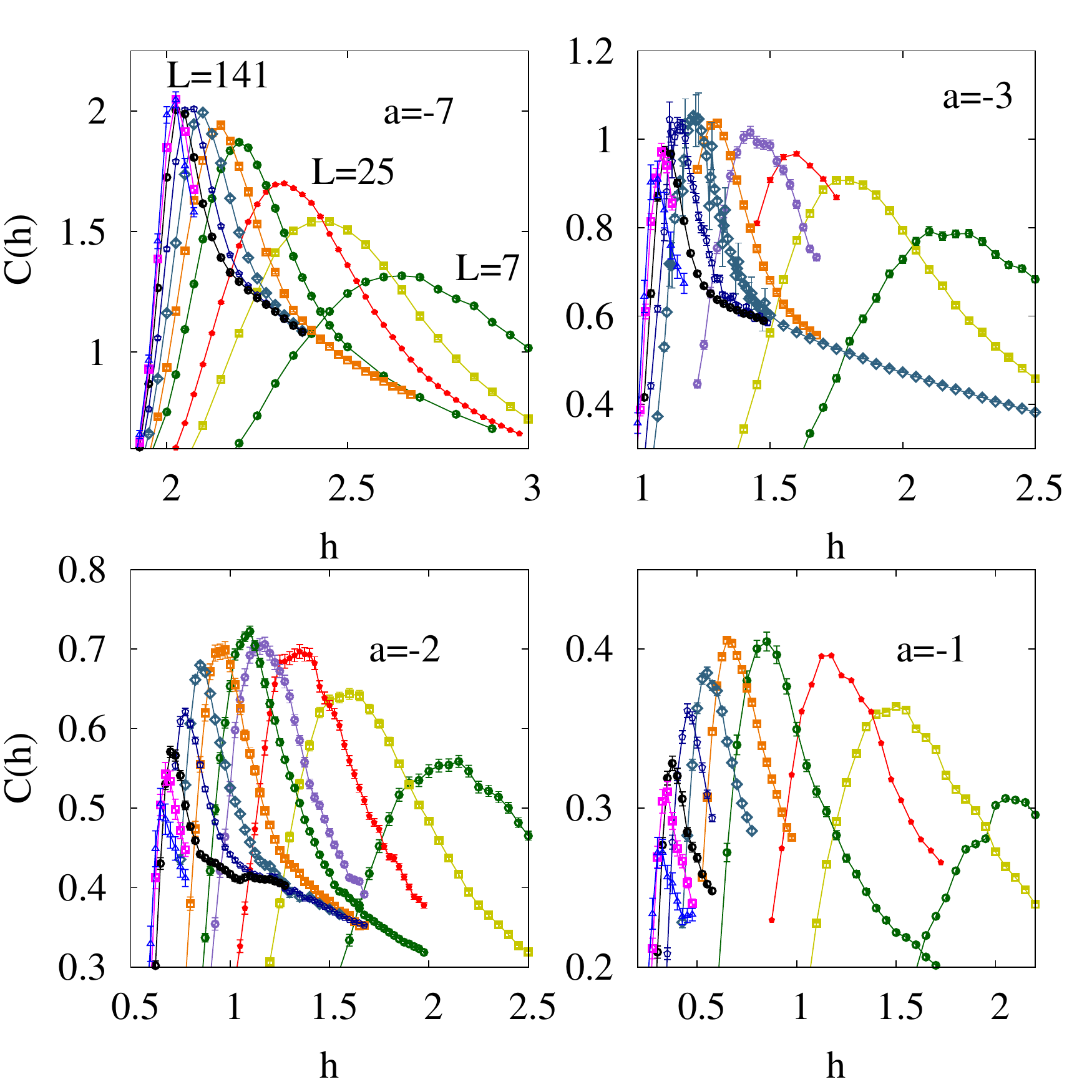}
\caption{(color online) Specific-heat-like quantity for different correlation strengths 
$a=-7\dots a=-1$ and system sizes $L=7, 11, 15, 21, 25, 35,
49, 69, 97, 117$ and $141$. For all values of $a$ the curves appear in monotonic ordering, meaning the curve of $L=7$ is on the very right and $L=141$ is on the very left, as labeled on the upper left subplot. The lines are guides to the eyes only.\label{allspecH}}
\end{center}
\end{figure}
Since the Binder cumulant exhibits no clear crossing (see
Fig.\ \ref{fig:binder}) one could suspect that
no phase transition is present. This is not the case as well
will see in the following.
 To determine the phase transition points, we start by
considering the average specific heat. In Fig.\ \ref{allspecH} the
results can be seen for $a=-7$ to $a=-1$. 
As for the uncorrelated RFIM, 
peaks can be observed clearly, which give evidence for
the existence of a phase transition also for the correlated case.
 We estimate the peaks by fitting parabolas over different intervals close to the maximum (for
every bootstrap sample).  The   positions of the peaks
in Fig.\ \ref{allspecH} move from right to left for increasing system
sizes.  To obtain the infinite-size limiting value $h_c$ and
an estimate for the critical exponent $\nu$ of the correlation
length, we fit the
positions of the peaks to Eq.\ (\ref{finite_hin_f}), resulting in fit
values as shown in the  upper part of Tab.\ \ref{results_tab_2}. 
Note that when determining the error bars from model
fitting, we usually have not only
taken the statistical error obtained from the fit routine (of the
{\tt gnuplot} program) but
we have always also varied the range of sizes, to get an
impression of possible systematic errors.

\begin{figure}[tph]
\begin{center}
\includegraphics[width=0.5\textwidth]{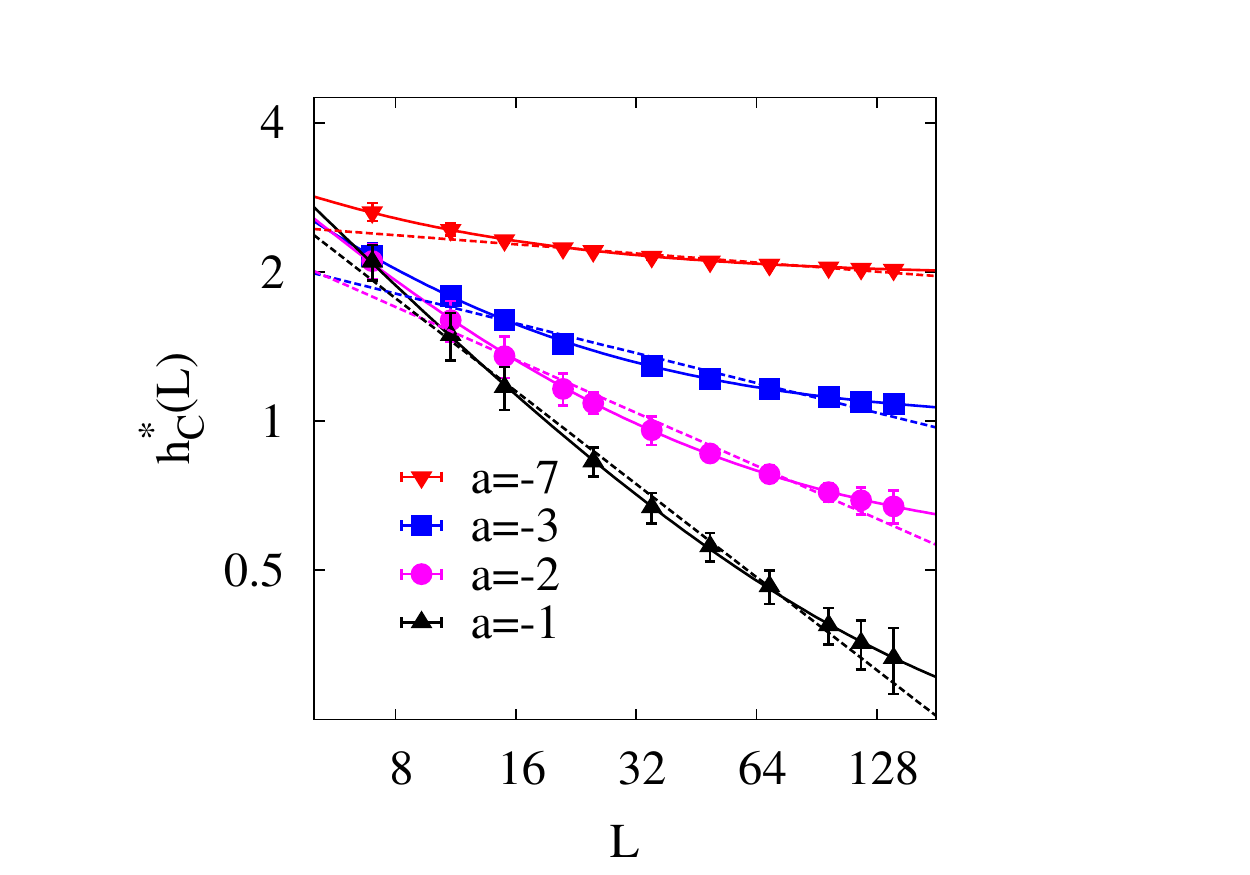}
\caption{(color online) 
Peak positions of the specific-heat-like quantity as function of the system sizes for $a=-7\dots a=-1$. The solid lines are fits assuming a finite critical value according to Eq.\ \ref{finite_hin_f}, the broken ones are fits imply $h_c=0$. \label{allspecHpeakpos}}
\end{center}
\end{figure}

In  Fig.\ \ref{allspecH} for the case  $a=-1$, 
the peaks move very close to $h=0$ and the
result from the fit for $h_c$ is also close to zero. Therefore,
another sensible ansatz is to set $h_c=0$. Fit parameters for this ansatz
are also shown in Tab.\ \ref{results_tab_2} in the lower part. Both
models are plotted in Fig.\ \ref{allspecHpeakpos} as solid or broken
lines, respectively. For $a<-1$,
 clearly $h_c>0$ is better compatible with the data points, while
for $a=-1$ no real decision can be made here between $h_c>0$ and $h_c=0$. 
Nevertheless, $h_c(L)$ also looks slightly curved in the double-logarithmic
plot, hence $h_c>0$ appears more likely here as well.

\begin{table}[th]
\begin{tabular}{|l | r | r | r| r |}\hline
							&$a=-7$ 	& $a=-3$ 		&$a=-2$ 	&$a=-1$ \\ \hline
$h_c$					&$1.962(4)$	&$0.992(8)$		&$0.537(8)$	&$0.183(9)$\\ \hline
$f$							&$3.2(1)$	&$6.0(3)$		&$7.5(2)$	&$9.8(3)$\\ \hline
$1/\nu$					& $0.79(2)$ 	& $0.84(2)$		&$0.81(2)$	&$0.84(2)$\\ \hline
\multicolumn{5}{l}{assuming no critical point $\Leftrightarrow h_c=0$}\\\hline
$f_0$						&$2.62(5)$	&$2.61(8)$		&$3.1(1)$	&$5.5(3)$\\ \hline
$1/\nu_0$				&$0.055(5)$	&$0.187(8)$		&$0.32(1)$	&$0.58(1)$\\ \hline
\end{tabular}
\caption{Fit parameters of Eq.\ \ref{finite_hin_f} for the peak position of the specific-heat-like quantity. The upper part contains the parameters for finite $h_c$. For the lower part $h_c=0$ was fixed. \label{results_tab_2}} 
\end{table}

\begin{figure}[tph]
\begin{center}
\includegraphics[width=0.5\textwidth]{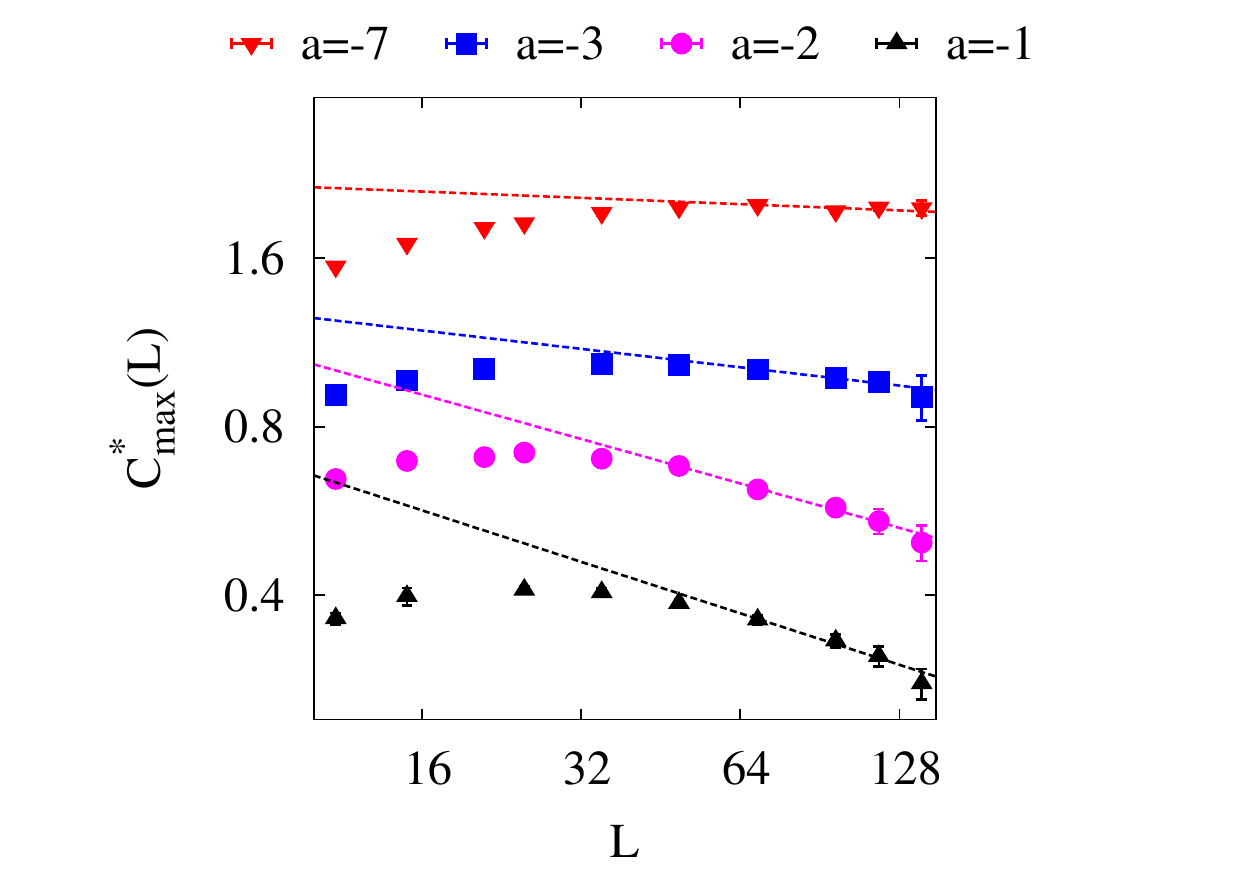}
\caption{(color online) Peak heights of the specific heat as function
of the system sizes for $a=-7\dots a=-1$. The lines are fits assuming
a  power law decay for large system sizes. 
\label{allspecHpeakheights}}
\end{center}
\end{figure}

To determine the critical exponent $\alpha$ according to Eq.\
(\ref{specHRescale}), we analyzed the peak heights of the specific
heat as shown in Fig.\ \ref{allspecHpeakheights}. They increase up to
$L \approx 50$ for all correlation strengths $a$ and decrease for
larger $L$. Thus, no clear scaling is visible. This could
be due to very strong finite-size corrections. Therefore,
under the assumption that the specific heat decreases in a power-law
fashion, we fitted the data points for
 very large system sizes a power law of the form
\begin{align}
C(h,L) &=  kL^{\alpha/\nu}\,. \label{eq_sh_peak_height}
\end{align}
The achieved exponents are small and negative can be
found in Tab. \ref{results_tab_C_max}.On the other
hand it may be that the specific heat levels off for even larger
system sizes, which would give the leading behavior $\alpha=0$.
In Sec.\ \ref{Con}, we will discuss these two options
in connection with the Rushbrooke inequality \cite{rushbrooke1963}
 and see that
$\alpha=0$ appears to be more likely.

\begin{table}[bh]
\begin{tabular}{|l | r | r | r| r |}\hline
			&$a=-7$ 	& $a=-3$ 		&$a=-2$ 	&$a=-1$ \\ \hline
$k$			&$2.3(2)$	&$1.7(1)$		&$1.9(1)$	&$1.0(1)$\\ \hline
$\alpha/\nu$	& $-0.04(2)$ 	& $-0.11(2)$		&$-0.26(1)$	&$-0.24(3)$\\ \hline
\end{tabular}
\caption{Fit parameters of Eq.\ \ref{eq_sh_peak_height} of the
 height of the maxima of the specific-heat as shown in 
Fig.\ \ref{allspecHpeakheights} 
\label{results_tab_C_max}} 
\end{table}

 \begin{figure}[tph]
\begin{center}
\includegraphics[width=0.5\textwidth]{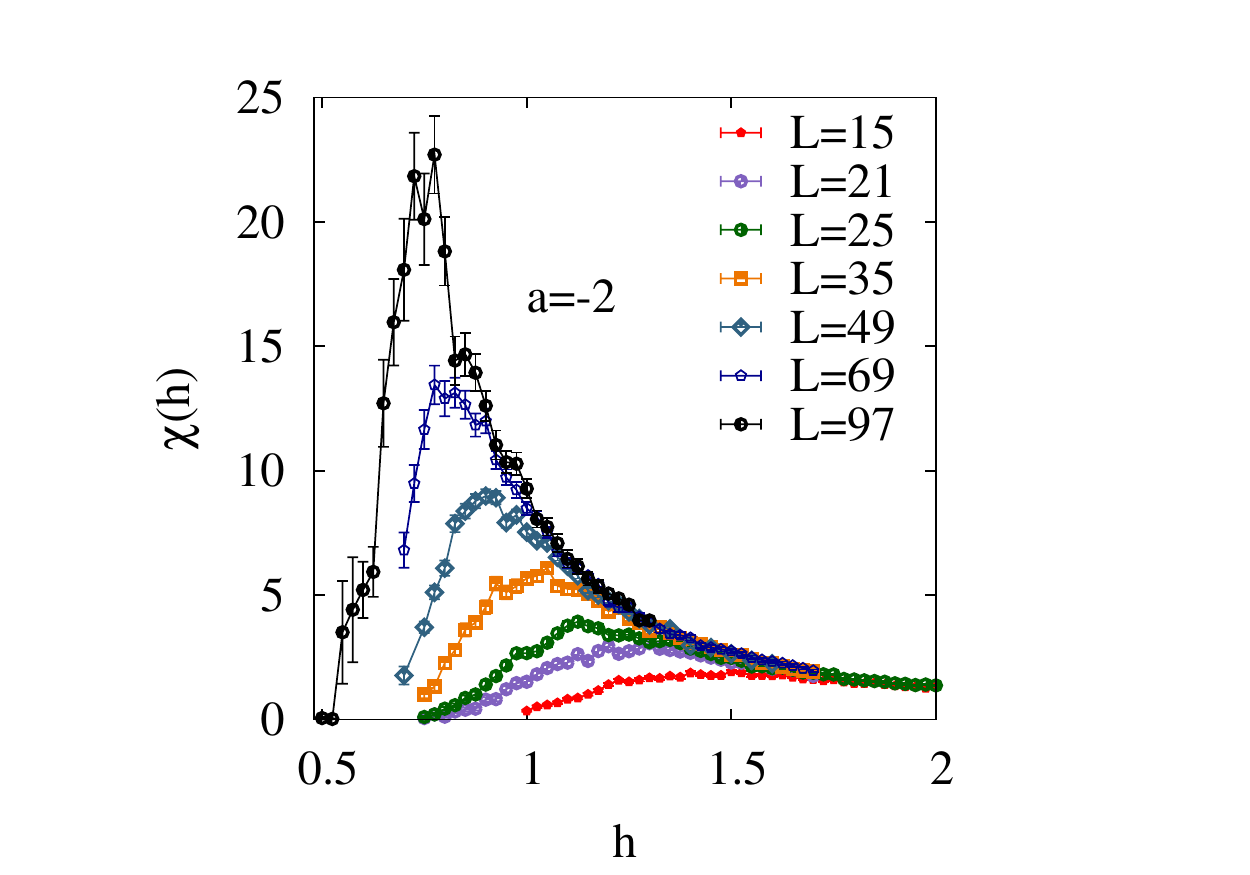}
\caption{(color online) Susceptibility for $a=-2$.\label{a2_susz} }
\end{center}
\end{figure}

We now turn to the susceptibility. The phase transition is
signaled by  a divergence of the susceptibility. An increasing peak
can be seen for $ a=-2 $ in  Fig.\ \ref{a2_susz} as example.   The
peaks are estimated in the same way as for the specific heat. The
resulting maxima are tuples $\left(h^*(L),\chi_{max}(L)\right)$ of
position and height.

 \begin{figure}[tph]
\begin{center}
\includegraphics[width=0.5\textwidth]{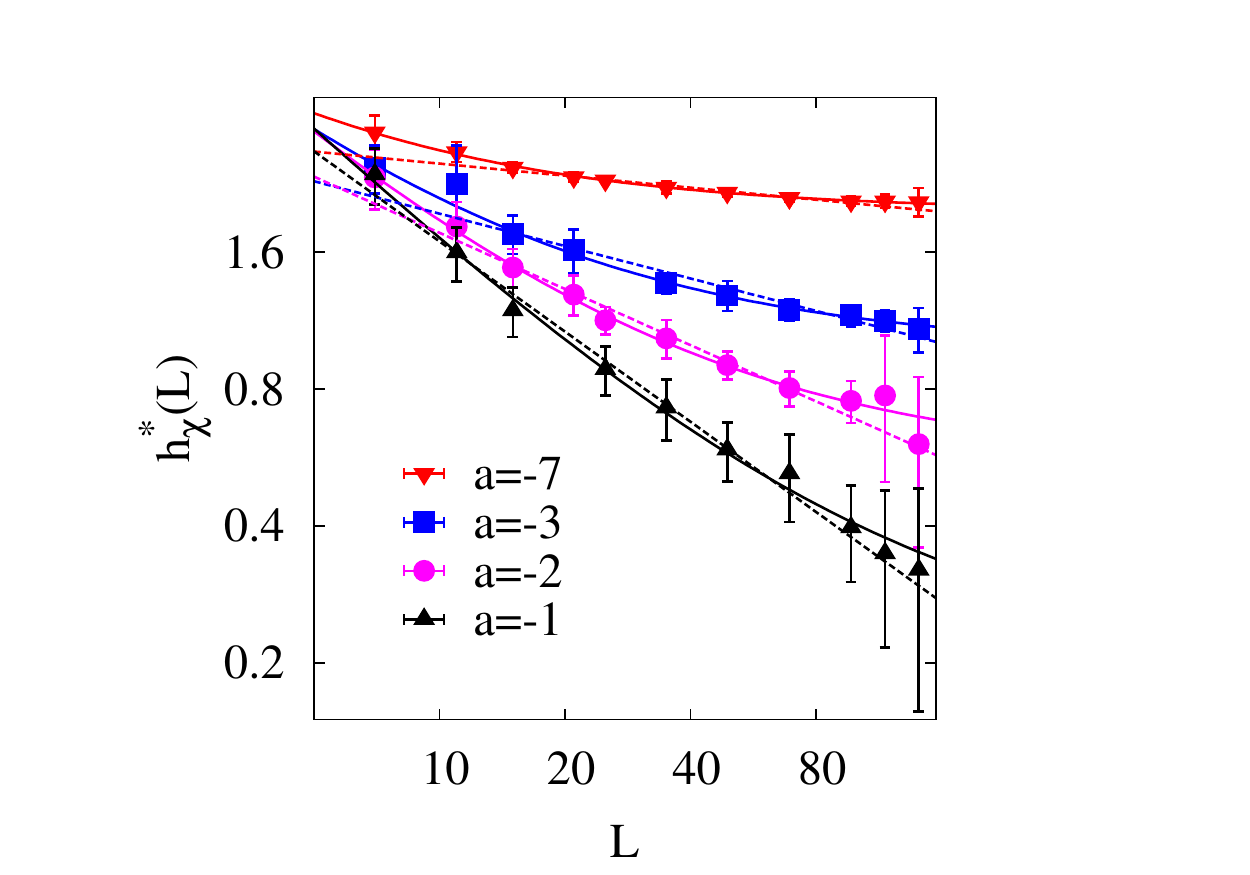}
\caption{(color online) Peak positions of the susceptibility as
function of the system sizes for $a=-7\dots a=-1$. The solid lines are
fits assuming a finite critical value according to Eq.\
\ref{finite_hin_f}, the broken ones are fits imply
$h_c=0$. \label{allsuszpeakpos}. }
\end{center}
\end{figure}

\begin{figure}[tph]
\begin{center}
\hspace{-2em}\includegraphics[width=0.4\textwidth]{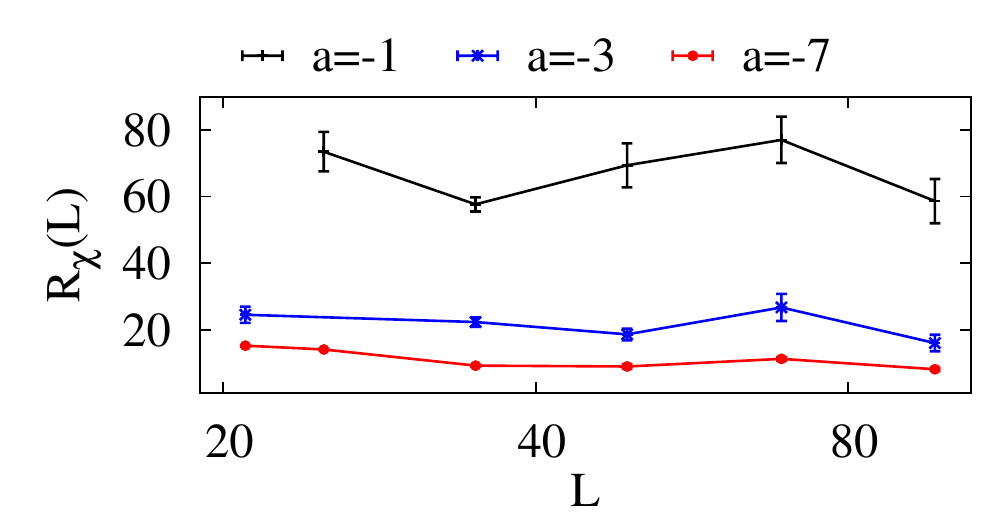}
\caption{(color online) Ratio $R_\chi\!(L)$ for $a=-7,-3-1$. The data points for $a=-2$ lie between $a=-7$ and $a=-3$ and are not included for
better visibility.\label{LoSA} }
\end{center}
\end{figure}

For the peak position of the susceptibility we assumed the same model
as we did for the specific heat.  The models and data points
can be found in Fig.\ \ref{allsuszpeakpos}.In particular
for $a=-1$, the error bars are quite large, despite the large number of 
samples, which is for the largest system sizes considerably higher
compared to the cases $a<-1$.
To understand this behavior we studied the degree of non-self averaging
 \cite{WisemanDomany1998} and we calculated
\begin{align}
R_\chi\!(L)=\text{var}(\chi(L))/\langle \chi(L)\rangle^2.
\end{align}
We found $R_\chi$ to stay approximately constant for increasing $L$, 
as shown in Fig.\ \ref{LoSA} for susceptibility measured at
the peak positions. We found
the same behavior qualitatively for different fixed values of $h$,
which shows that the correlated
RFIM is non-self averaging for a large range of the disorder parameter,
as many other systems exhibiting quenched disorder.
In particular, the results in  Fig.\ \ref{LoSA} show that the
degree of non-self averaging is strongest for $a=-1$, which explains
the large error bars. To achieve much smaller error bars for
the susceptibility, a much larger number of samples would be necessary,
which is beyond the capacity of our numerical resources.

For the finite-size scaling
of the peak-positions, we tested Eq.\ (\ref{finite_hin_f}) using 
the saturating
ansatz ($h_c$ included in the fit) 
as well as a pure power law decay (via $h_c\equiv 0$). 
 The fit parameters for
both models can be found in Tab.\ \ref{results_tab_susz_peak_pos}.
Again the saturating
model brings up, within the present accuracy,  the same infinite-size
critical point $h_c$ as we found before.
Due to the error bars, we can not rule out a $h_c=0$ for $a=-1$, while for $a<-1$ again clearly
$h_c>0$ holds.

\begin{table}[bph]
\begin{tabular}{|l | r | r| r | r |}\hline
						& $a=-7$		& $a=-3$ 		&$a=-2$ 	&$a=-1$ 	\\ \hline
$h_c$				&$1.97(1)$		&$0.98(4)$		&$0.56(4)$	&$0.20(5)$	\\ \hline
$f$						&$4.8(5)$		&$7(1)$			&$9(1)$		&$11(2)$	\\ \hline
$1/\nu$				&$0.83(5)$		&$0.84(8)$		&$0.85(7)$	&$0.88(8)$	\\ \hline
\multicolumn{5}{l}{assuming no critical point $\Leftrightarrow h_c=0$}\\\hline
$f_0$					&$3.06(8)$		&$3.4(3)$		&$4.5(4)$	&$7.7(7)$\\ \hline
$1/\nu_0$			&$0.088(7)$		&$0.24(2)$		&$0.41(3)$	&$0.66(3)$\\ \hline
\end{tabular}
\caption{Fit parameters of Eq.\ \ref{finite_hin_f} of the peak position of the susceptibility. The upper part contains the parameters for finite $h_c$. For the lower part $h_c=0$ was fixed. \label{results_tab_susz_peak_pos}} 
\end{table}

\begin{table}[t!h]
\begin{tabular}{|l | r | r | r | r |}\hline
						&$a=-7$		& $a=-3$ 		&$a=-2$ 	&$a=-1$ \\ \hline
$b_0$					&$ 0.064(3)$	&$0.049(2)$		&$0.049(2)$	&$0.041(3)$\\ \hline
$\gamma/\nu$	& $1.56(1) $ 	& $1.45(1)$		&$1.34(2)$	&$1.20(2)$\\ \hline
\end{tabular}
\caption{Fit parameters of the peak heights of the susceptibility when fitting according to $b_0L^{\gamma/\nu}$.
 \label{results_tab_susz_height}} 
\end{table}

 \begin{figure}[tph]
\begin{center}
\includegraphics[width=0.5\textwidth]{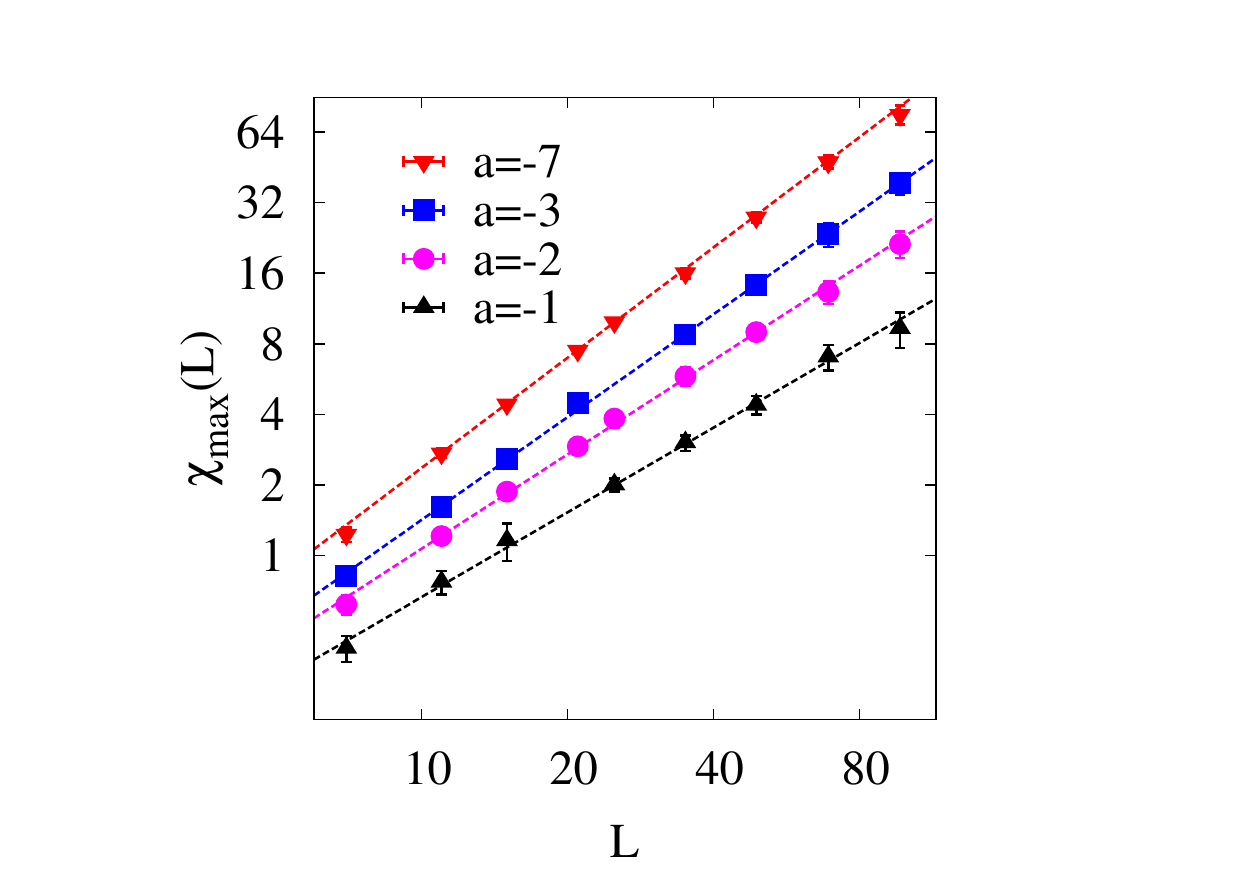}
\caption{(color online) Peak heights of the susceptibility as function 
of the system sizes for $a=-3\dots a=-1$. 
\label{allsuszpeakheights}}
\end{center}
\end{figure}

In contrast to the specific heat, the peak
heights of the susceptibility shows a clear power law behaviour for all studied correlation
strengths, see Fig.\ \ref{allsuszpeakheights}. 
Thus, in the thermodynamic limit the susceptibility diverges.
Compared to the peak positions displayed in Fig.\ \ref{allsuszpeakpos}, the
fluctuations for the peak height are much smaller, thus a clear
power-law behavior is visible. The critical exponents $\gamma/\nu$, as obtained
from a power-law fit, are displayed in Tab.\ \ref{results_tab_susz_height}.
The values  decreases with increasing $a$. 

\begin{figure}[th]
\begin{center}
\includegraphics[width=0.5\textwidth]{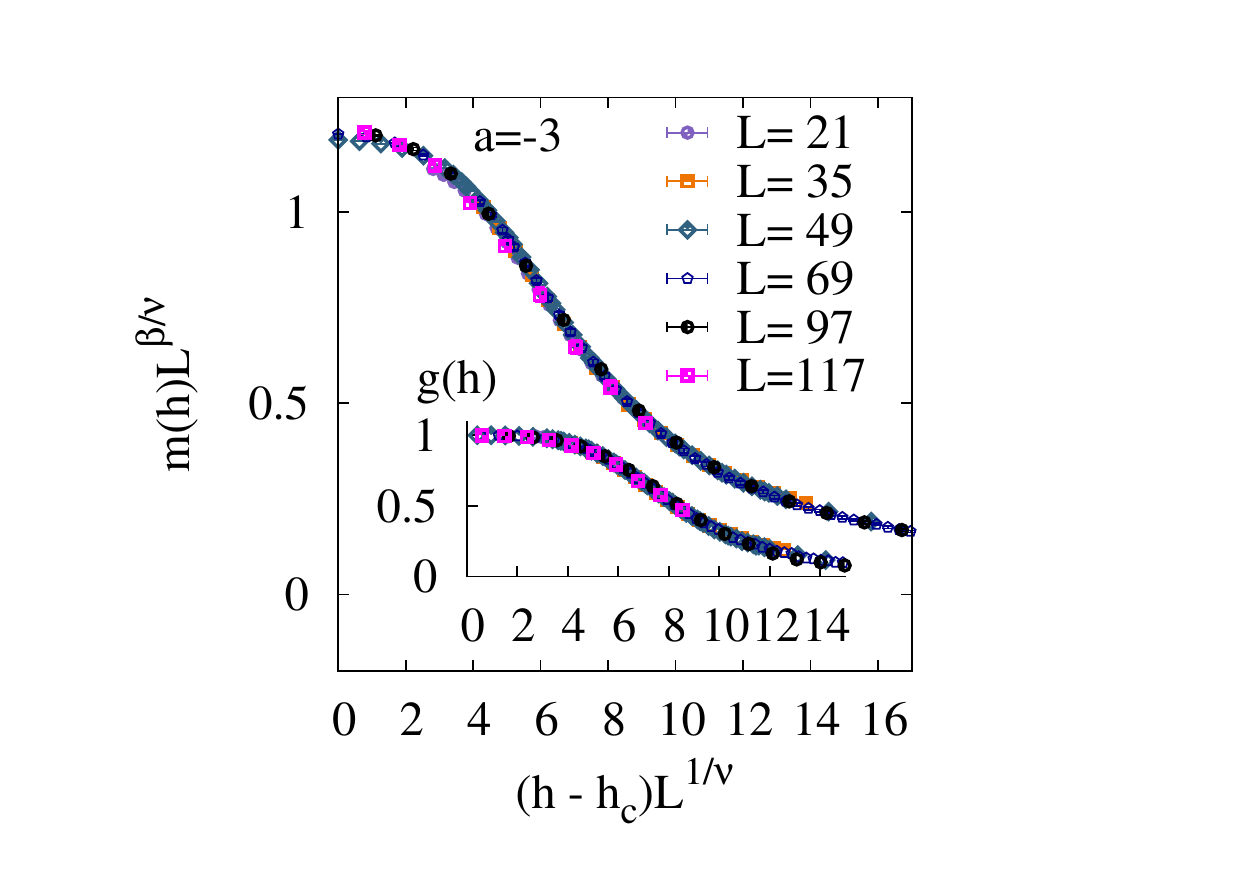}
\caption{(color online) 
Data collapse of the magnetization and of the Binder cumulant (inset) 
for $a=-3$.
\label{binder_collapse_a-3_inset_magn_a-3}}
\end{center}
\end{figure}

For a finite-size analysis of the Binder cumulant and of the
magnetization we performed data collapses according
to Eqs.\ (\ref{binderRescale}) and (\ref{magRescale}). Example
data collapses for the magnetization and as inset for the
Binder cumulant for $a=-3$ are shown in Fig.\
\ref{binder_collapse_a-3_inset_magn_a-3}. 
One sees that the quality of the collapses is very good.
These lead to sets of critical values and
exponents as shown in Tab.\ \ref{results_tab_1}.

\begin{table}[th]
\begin{tabular}{|l| l | l | l| l |}\hline
				& $a=-7$& $a=-3$ 		&$a=-2$ 	&$a=-1$ \\ \hline
$h_c$ 	 					&$1.94(1)$	& $0.95(2)$		&$0.47(4)$	&$0.12(3)$\\ \hline
$1/\nu$						&$0.78(2)$	& $0.77(4)$		&$0.75(3)$	&$0.78(4)$\\ \hline
$\beta/\nu$					&$0.005(5)$	& $0.03(2)$		&$0.01(1)$	&$0.01(8)$\\ \hline
\end{tabular}
\caption{Critical value $h_c$ and correlations length exponent $\nu$ derived from the finite size scaling analysis of the Binder cumulant and the magnetization. These values are obtained via data collapses.
\label{results_tab_1}} 
\end{table}
\section{\label{Arg} Minimum and maximum correlated disorder}
\begin{figure}[hb]
\begin{center}
\includegraphics[width=0.25\textwidth, angle=270]{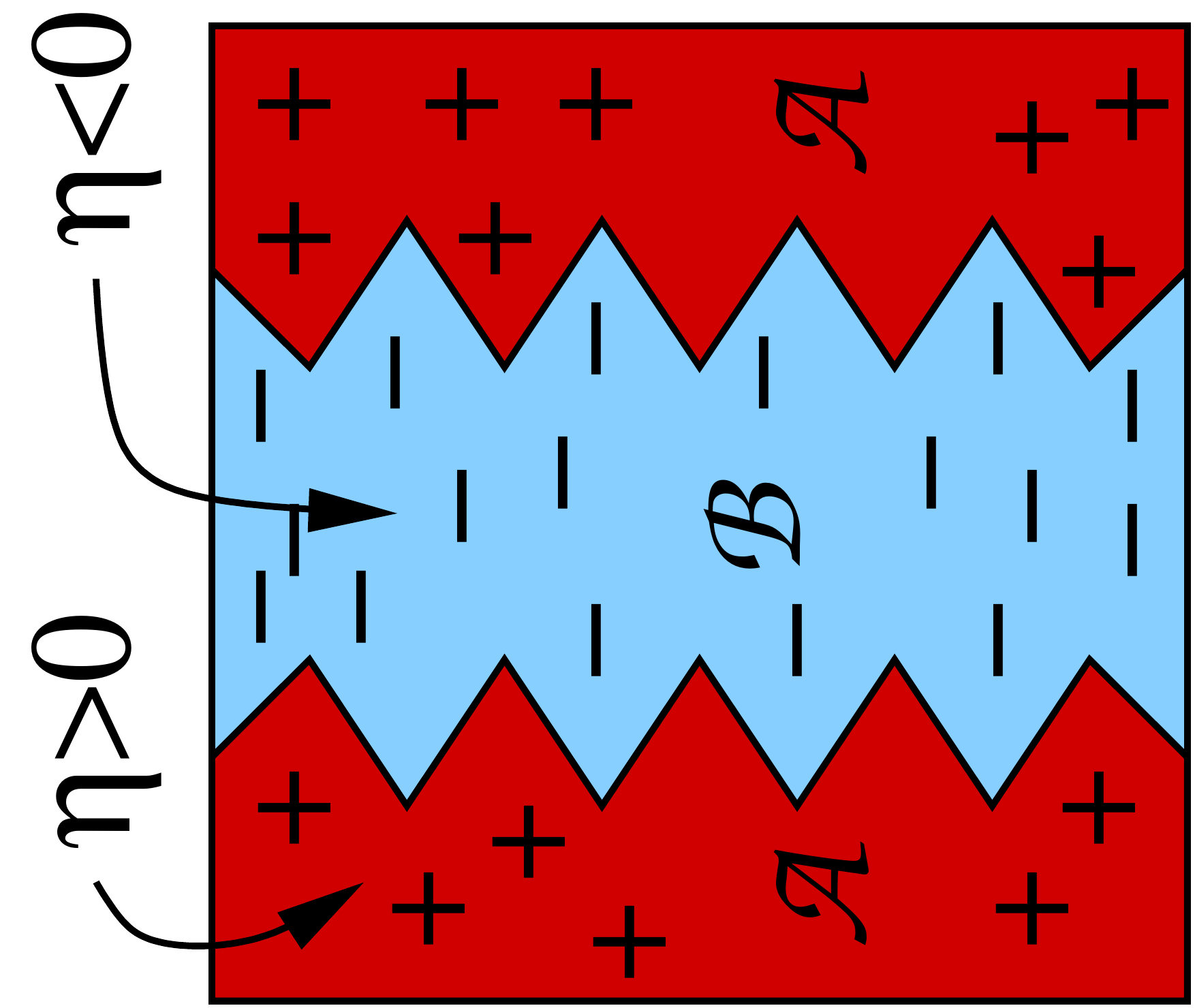}
\caption{(color online) System with bi-parted random field (with periodic boundaries).
 \label{domain_wall}}
\end{center}
\end{figure}

The minimum correlation range of the disorder is $a\rightarrow -\infty \equiv \delta$-correlated disorder, i.e. the normal RFIM.
The other extreme is $a\to 0^-$. As it is illustrated in Fig.\ \ref{3corrExample}, by increasing the correlation strength $a$, the regions of sites with almost the same sign of the field get larger and larger, while keeping
the same average close to 0. Hence, for $a\to0^-$ one can imagine
each realization of the disorder being bi-parted. Bi-parted
 means, to find two distinct clusters
$\mathcal{A}=\{\eta_i>0\}$ and $\mathcal{B}=\{\eta_i<0\}$, see Fig.\
\ref{domain_wall}. An Imry-Ma type of argument would read as follows:
In such a  disorder realization, for a state where all spins are aligned with its local field, the resulting interface energy
between the clusters $\mathcal{A,B}$ would be $E_I\sim L^{d_f}$ 
with fractal exponent $d-1\le d_f\le d$. This
competes against the field energy $E_h\sim -hL^d$: If
$E_I < |E_h|$, the ground-state will be an ordered phase, otherwise both
clusters are locally aligned. A $T=0$ phase transition occurs when
$|E_h|=E_I$. From this we see that the
 finite-size critical point scales in the limit $a\to 0^-$ 
as  $h_c\sim L^{d_f-d}\to 0$ for $L\to\infty$. This means 
for highly correlated but arbitrarily small disorder the RFIM will 
behave as a super paramagnet in a field  in the thermodynamic limit. 
Though, there would be no disorder-driven phase transition anymore.
Furthermore, since the number of spins at distance
$r$ scales as $r^{d-1}$ but the disorder correlation decreases only 
as $r^{a}$, its appears plausible that even for a finite range of $a<0$ 
values the cumulative effect of the
correlation might dominate and indeed $h_c=0$ already for these values of $a$.
This explains why the result for $a=-1$ is ambiguous. 
 
\section{\label{Con}Conclusion and Discussion}
We have presented the results of exact ground-state calculations of the
RFIM with correlated disorder for different correlations strengths.
To numerically calculate the ground states, we have applied
a mapping to the maximum-flow problem. Using efficient
polynomial-time-running maximum-flow/minimum-cut algorithms,
we were able to study large systems sizes up to $N=141^3$.

We studied different quantities like magnetization, Binder cumulant,
susceptibility and a specific heat-like quantity and applied
finite-size scaling techniques to obtain the critical exponents.
The combined results for the critical exponents are shown in
Tab.\ \ref{resulttabs}.
 We tested the two possibilities for the
 values of $\alpha$, by applying the
Rushbrooke inequality $\alpha + 2\beta +\gamma \ge 2$ which
holds usually as equality.\cite{rushbrooke1965}
When choosing $\alpha=0$, the Rushbrooke equation
is fulfilled
in all cases within error bars. For the values of $\alpha$ quoted in
Tab.\ \ref{results_tab_C_max}, obtained via fitting the data for just the
few largest systems sizes, the Rushbrooke sum is
(assuming $h_c>0$) considerably smaller than 2  
for $a=-2,-1$. Hence, the value $\alpha=0$
appears to be more likely.
Note in all cases, the values quoted in the table 
are compatible within error bars with the results for the uncorrelated case, in particular due
to the relative large error bar for the critical exponent $\gamma$.
 Nevertheless, the data for the peak heights of the susceptibility 
(Fig.\ \ref{allsuszpeakheights}) show a trend towards a smaller slope
when increasing $a$ from $-7$ to $-1$: The results for $\gamma/\nu$, which
are very precise (see Tab.\ \ref{results_tab_susz_height}) are clearly different within error bars.
In this case, to still fulfill the Rushbrooke inequality, the true
value for $\nu$, in particular for  values $a=-2$ and $a=-1$,
 should be larger, at or somehow above the upper bounds
given the standard error bars. Hence, it is quite likely that the correlation of the disorder
creates non-universality for the RFIM, as in the case of the
diluted ferromagnet.\cite{BallesterosParisi1999}.
Also, among our results, the case $a=-1$ is special. 
As discussed above, it appears plausible for $a=-1$ that the cumulative 
effect of the correlation might dominate. This would lead to $h_c=0$ 
already for $a=-1$.
The result $h_c=0$ is indeed possible, as shown in last row of 
Tab.\ \ref{resulttabs}:
Also on the base of the Rushbrooke sum, we cannot take  
decision on this issue. Nevertheless, 
numerically the inequality is much better fulfilled when assuming $h_c=0$.

\begin{table}[bph]
\begin{center}
 \begin{tabular}{|c|c|c|c|c|c|c|}\hline
$a$	  &  $h_c$ & $\alpha$	& $\beta$		& $\gamma$& $\nu$&  
RS sum\\ \hline
$-\infty$  & $2.27(1)$ & $\sim 0$ & $0.016(7)$ & $2.1(1)$ & $1.37(9)$ & 
$2.1(2)$ \\ \hline
$-7  $& $1.96(2)$ & $0$ 		&$  	0.01(1)		$&$	2.0(3)  $&$ 1.26(8)$&$ 2.0(3)$\\ \hline
$-3  $ & $0.97(2)$ &$0$		&$   0.04(6)		$&$	1.8(4)  $&$ 1.2(2)	 $&$ 1.8(5)$ \\ \hline
$-2  $&  $0.52(5)$ & $0$		&$ 	0.01(3)		$&$	1.7(3)   $&$1.2(2)  $&$ 1.7(4)$ \\ \hline
$-1 $&  $ 0.17(5)$ & $0$		&$ 	0.05(10)		$&$	1.5(3)   $&$1.2(2) $&$  1.5(5)$ \\ \hline
$-1^*$ &$0$ & $0$ 		&$   0.1(1)		$&$ 1.9(1)   $&$1.61(9)$&$ 2.0(4)$ \\ \hline
 \end{tabular}
\caption{Final results for the value of $h_c$, the 
critical exponents $\alpha,\beta,\gamma$ and $\nu$ and
the Rushbrooke (RS) sum $\alpha+2\beta+\gamma$. The first line
shows the result for the uncorrelated case taken from 
Refs.\ \onlinecite{HartmannYoung2001,middleton2002}.
The last line is for the assumption $h_c(a=-1)=0$.
\label{resulttabs}
}
\end{center}
\end{table}

%

%
%
%
\section{\label{Ack}Acknowledgments}
We would like to thank K. Janzen  for fruitful discussions and critically
reading the manuscript. Furthermore, 
we are grateful to M. Niemann helpful comments.
The calculations were carried out on \textbf{GOLEM} (\textbf{G}ro\ss rechner \textbf{OL}denburg f\"ur \textbf{E}xplizit \textbf{M}ultidisziplin\"are \textbf{F}orschung) and the \textbf{HERO} (\textbf{H}igh-\textbf{E}nd Computing \textbf{R}esource \textbf{O}ldenburg) at the University of Oldenburg.
\newpage
\bibliographystyle{apsrev}
\bibliography{Literatur}

\begin{thebibliography}{40}
\expandafter\ifx\csname natexlab\endcsname\relax\def\natexlab#1{#1}\fi
\expandafter\ifx\csname bibnamefont\endcsname\relax
  \def\bibnamefont#1{#1}\fi
\expandafter\ifx\csname bibfnamefont\endcsname\relax
  \def\bibfnamefont#1{#1}\fi
\expandafter\ifx\csname citenamefont\endcsname\relax
  \def\citenamefont#1{#1}\fi
\expandafter\ifx\csname url\endcsname\relax
  \def\url#1{\texttt{#1}}\fi
\expandafter\ifx\csname urlprefix\endcsname\relax\def\urlprefix{URL }\fi
\providecommand{\bibinfo}[2]{#2}
\providecommand{\eprint}[2][]{\url{#2}}

\bibitem[{\citenamefont{Bricomont and Kupiainen}(1987)}]{bricmont1987}
\bibinfo{author}{\bibfnamefont{J.}~\bibnamefont{Bricomont}} \bibnamefont{and}
  \bibinfo{author}{\bibfnamefont{A.}~\bibnamefont{Kupiainen}},
  \bibinfo{journal}{Phys. Rev. Lett.} \textbf{\bibinfo{volume}{59}},
  \bibinfo{pages}{1829} (\bibinfo{year}{1987}).

\bibitem[{\citenamefont{Gofman et~al.}(1993)\citenamefont{Gofman, Adler,
  Aharony, Harris, and Schwartz}}]{GofmanAdlerAharonyHarrisSchwartz1993}
\bibinfo{author}{\bibfnamefont{M.}~\bibnamefont{Gofman}},
  \bibinfo{author}{\bibfnamefont{J.}~\bibnamefont{Adler}},
  \bibinfo{author}{\bibfnamefont{A.}~\bibnamefont{Aharony}},
  \bibinfo{author}{\bibfnamefont{A.~B.} \bibnamefont{Harris}},
  \bibnamefont{and} \bibinfo{author}{\bibfnamefont{M.}~\bibnamefont{Schwartz}},
  \bibinfo{journal}{Phys. Rev. Lett.} \textbf{\bibinfo{volume}{71}},
  \bibinfo{pages}{1569} (\bibinfo{year}{1993}).

\bibitem[{\citenamefont{Rieger}(1995)}]{Rieger1995}
\bibinfo{author}{\bibfnamefont{H.}~\bibnamefont{Rieger}},
  \bibinfo{journal}{Phys. Rev. B} \textbf{\bibinfo{volume}{52}},
  \bibinfo{pages}{6659} (\bibinfo{year}{1995}).

\bibitem[{\citenamefont{Nowak et~al.}(1998)\citenamefont{Nowak, Usadel, and
  Esser}}]{Nowak1998}
\bibinfo{author}{\bibfnamefont{U.}~\bibnamefont{Nowak}},
  \bibinfo{author}{\bibfnamefont{K.~D.} \bibnamefont{Usadel}},
  \bibnamefont{and} \bibinfo{author}{\bibfnamefont{J.}~\bibnamefont{Esser}},
  \bibinfo{journal}{Physica A: Statistical and Theoretical Physics}
  \textbf{\bibinfo{volume}{250}}, \bibinfo{pages}{1 } (\bibinfo{year}{1998}).

\bibitem[{\citenamefont{Hartmann and Nowak}(1999)}]{art_uli1999}
\bibinfo{author}{\bibfnamefont{A.~K.} \bibnamefont{Hartmann}} \bibnamefont{and}
  \bibinfo{author}{\bibfnamefont{U.}~\bibnamefont{Nowak}},
  \bibinfo{journal}{Eur. Phys. J. B} \textbf{\bibinfo{volume}{7}},
  \bibinfo{pages}{105} (\bibinfo{year}{1999}).

\bibitem[{\citenamefont{Hartmann and Young}(2001)}]{HartmannYoung2001}
\bibinfo{author}{\bibfnamefont{A.~K.} \bibnamefont{Hartmann}} \bibnamefont{and}
  \bibinfo{author}{\bibfnamefont{A.~P.} \bibnamefont{Young}},
  \bibinfo{journal}{Phys. Rev. B} \textbf{\bibinfo{volume}{64}},
  \bibinfo{pages}{214419} (\bibinfo{year}{2001}).

\bibitem[{\citenamefont{Middleton and Fisher}(2002)}]{middleton2002}
\bibinfo{author}{\bibfnamefont{A.~A.} \bibnamefont{Middleton}}
  \bibnamefont{and} \bibinfo{author}{\bibfnamefont{D.~S.}
  \bibnamefont{Fisher}}, \bibinfo{journal}{Phys. Rev. B}
  \textbf{\bibinfo{volume}{65}}, \bibinfo{pages}{134411}
  (\bibinfo{year}{2002}).

\bibitem[{\citenamefont{Frontera and Vives}(2002)}]{frontera2002}
\bibinfo{author}{\bibfnamefont{C.}~\bibnamefont{Frontera}} \bibnamefont{and}
  \bibinfo{author}{\bibfnamefont{E.}~\bibnamefont{Vives}},
  \bibinfo{journal}{Computer Physics Communications}
  \textbf{\bibinfo{volume}{147}}, \bibinfo{pages}{455 } (\bibinfo{year}{2002}).

\bibitem[{\citenamefont{Sepp\"al\"a et~al.}(2002)\citenamefont{Sepp\"al\"a,
  Pulkkinen, and Alava}}]{seppala2002}
\bibinfo{author}{\bibfnamefont{E.~T.} \bibnamefont{Sepp\"al\"a}},
  \bibinfo{author}{\bibfnamefont{A.~M.} \bibnamefont{Pulkkinen}},
  \bibnamefont{and} \bibinfo{author}{\bibfnamefont{M.~J.} \bibnamefont{Alava}},
  \bibinfo{journal}{Phys. Rev. B} \textbf{\bibinfo{volume}{66}},
  \bibinfo{pages}{144403} (\bibinfo{year}{2002}).

\bibitem[{\citenamefont{Hartmann}(2002)}]{Hartmann2002}
\bibinfo{author}{\bibfnamefont{A.~K.} \bibnamefont{Hartmann}},
  \bibinfo{journal}{Phys. Rev. B} \textbf{\bibinfo{volume}{65}},
  \bibinfo{pages}{174427} (\bibinfo{year}{2002}).

\bibitem[{\citenamefont{Middleton}(2002)}]{middleton2002b}
\bibinfo{author}{\bibfnamefont{A.~A.} \bibnamefont{Middleton}},
  \bibinfo{journal}{preprint arXiv:cond-mat/0208182}  (\bibinfo{year}{2002}).

\bibitem[{\citenamefont{Zumsande et~al.}(2008)\citenamefont{Zumsande, Alava,
  and Hartmann}}]{fes_rfim2008}
\bibinfo{author}{\bibfnamefont{M.}~\bibnamefont{Zumsande}},
  \bibinfo{author}{\bibfnamefont{M.~J.} \bibnamefont{Alava}}, \bibnamefont{and}
  \bibinfo{author}{\bibfnamefont{A.~K.} \bibnamefont{Hartmann}},
  \bibinfo{journal}{J. Stat. Mech.} p. \bibinfo{pages}{P02012}
  (\bibinfo{year}{2008}).

\bibitem[{\citenamefont{Ahrens and Hartmann}(2011)}]{AhrensHartmann2011}
\bibinfo{author}{\bibfnamefont{B.}~\bibnamefont{Ahrens}} \bibnamefont{and}
  \bibinfo{author}{\bibfnamefont{A.~K.} \bibnamefont{Hartmann}},
  \bibinfo{journal}{Phys. Rev. B} \textbf{\bibinfo{volume}{83}},
  \bibinfo{pages}{014205} (\bibinfo{year}{2011}).

\bibitem[{\citenamefont{Fedorenko and K\"uhnel}(2007)}]{fedorenko2007}
\bibinfo{author}{\bibfnamefont{A.~A.} \bibnamefont{Fedorenko}}
  \bibnamefont{and} \bibinfo{author}{\bibfnamefont{F.}~\bibnamefont{K\"uhnel}},
  \bibinfo{journal}{Phys. Rev. B} \textbf{\bibinfo{volume}{75}},
  \bibinfo{pages}{174206} (\bibinfo{year}{2007}).

\bibitem[{\citenamefont{Makse et~al.}(1996)\citenamefont{Makse, Havlin,
  Schwartz, and Stanley}}]{MakseHavlin1996}
\bibinfo{author}{\bibfnamefont{H.~A.} \bibnamefont{Makse}},
  \bibinfo{author}{\bibfnamefont{S.}~\bibnamefont{Havlin}},
  \bibinfo{author}{\bibfnamefont{M.}~\bibnamefont{Schwartz}}, \bibnamefont{and}
  \bibinfo{author}{\bibfnamefont{H.~E.} \bibnamefont{Stanley}},
  \bibinfo{journal}{Phys. Rev. E} \textbf{\bibinfo{volume}{53}},
  \bibinfo{pages}{5445} (\bibinfo{year}{1996}).

\bibitem[{\citenamefont{Ballesteros and Parisi}(1999)}]{BallesterosParisi1999}
\bibinfo{author}{\bibfnamefont{H.~G.} \bibnamefont{Ballesteros}}
  \bibnamefont{and} \bibinfo{author}{\bibfnamefont{G.}~\bibnamefont{Parisi}},
  \bibinfo{journal}{Phys. Rev. B} \textbf{\bibinfo{volume}{60}},
  \bibinfo{pages}{12912} (\bibinfo{year}{1999}).

\bibitem[{\citenamefont{Hod and Keshet}(2004)}]{hod2004}
\bibinfo{author}{\bibfnamefont{S.}~\bibnamefont{Hod}} \bibnamefont{and}
  \bibinfo{author}{\bibfnamefont{U.}~\bibnamefont{Keshet}},
  \bibinfo{journal}{Phys. Rev. E} \textbf{\bibinfo{volume}{70}},
  \bibinfo{pages}{015104} (\bibinfo{year}{2004}).

\bibitem[{\citenamefont{Fedorenko}(2008)}]{fedorenko2008}
\bibinfo{author}{\bibfnamefont{A.~A.} \bibnamefont{Fedorenko}},
  \bibinfo{journal}{Phys. Rev. B} \textbf{\bibinfo{volume}{77}},
  \bibinfo{pages}{094203} (\bibinfo{year}{2008}).

\bibitem[{\citenamefont{Weinrib and Halperin}(1983)}]{WeinribHalperin1983}
\bibinfo{author}{\bibfnamefont{A.}~\bibnamefont{Weinrib}} \bibnamefont{and}
  \bibinfo{author}{\bibfnamefont{B.~I.} \bibnamefont{Halperin}},
  \bibinfo{journal}{Phys. Rev. B} \textbf{\bibinfo{volume}{27}},
  \bibinfo{pages}{413} (\bibinfo{year}{1983}).

\bibitem[{\citenamefont{Harris}(1974)}]{harris1974}
\bibinfo{author}{\bibfnamefont{A.~B.} \bibnamefont{Harris}},
  \bibinfo{journal}{Journal of Physics C: Solid State Physics}
  \textbf{\bibinfo{volume}{7}}, \bibinfo{pages}{1671} (\bibinfo{year}{1974}),
  \urlprefix\url{http://stacks.iop.org/0022-3719/7/i=9/a=009}.

\bibitem[{\citenamefont{Ballesteros et~al.}(1999)\citenamefont{Ballesteros,
  Fernández, Martín-Mayor, Sudupe, Parisi, and
  Ruiz-Lorenzo}}]{ballesteros1999}
\bibinfo{author}{\bibfnamefont{H.~G.} \bibnamefont{Ballesteros}},
  \bibinfo{author}{\bibfnamefont{L.~A.} \bibnamefont{Fernández}},
  \bibinfo{author}{\bibfnamefont{V.}~\bibnamefont{Martín-Mayor}},
  \bibinfo{author}{\bibfnamefont{A.~M.} \bibnamefont{Sudupe}},
  \bibinfo{author}{\bibfnamefont{G.}~\bibnamefont{Parisi}}, \bibnamefont{and}
  \bibinfo{author}{\bibfnamefont{J.~J.} \bibnamefont{Ruiz-Lorenzo}},
  \bibinfo{journal}{Journal of Physics A: Mathematical and General}
  \textbf{\bibinfo{volume}{32}}, \bibinfo{pages}{1} (\bibinfo{year}{1999}),
  \urlprefix\url{http://stacks.iop.org/0305-4470/32/i=1/a=004}.

\bibitem[{\citenamefont{Ballesteros et~al.}(1998)\citenamefont{Ballesteros,
  Fern\'andez, Mart\'\i{}n-Mayor, Mu\~noz Sudupe, Parisi, and
  Ruiz-Lorenzo}}]{ballesteros1998}
\bibinfo{author}{\bibfnamefont{H.~G.} \bibnamefont{Ballesteros}},
  \bibinfo{author}{\bibfnamefont{L.~A.} \bibnamefont{Fern\'andez}},
  \bibinfo{author}{\bibfnamefont{V.}~\bibnamefont{Mart\'\i{}n-Mayor}},
  \bibinfo{author}{\bibfnamefont{A.}~\bibnamefont{Mu\~noz Sudupe}},
  \bibinfo{author}{\bibfnamefont{G.}~\bibnamefont{Parisi}}, \bibnamefont{and}
  \bibinfo{author}{\bibfnamefont{J.~J.} \bibnamefont{Ruiz-Lorenzo}},
  \bibinfo{journal}{Phys. Rev. B} \textbf{\bibinfo{volume}{58}},
  \bibinfo{pages}{2740} (\bibinfo{year}{1998}).

\bibitem[{pap()}]{papercore_gen}
\bibinfo{note}{\emph{Papercore} is a free and open access database for
  summaries of scientific (currently mainly physics) papers.},
  \urlprefix\url{http://www.papercore.org/}.

\bibitem[{\citenamefont{Peng et~al.}(1991)\citenamefont{Peng, Havlin, Schwartz,
  and Stanley}}]{Peng1991}
\bibinfo{author}{\bibfnamefont{C.-K.} \bibnamefont{Peng}},
  \bibinfo{author}{\bibfnamefont{S.}~\bibnamefont{Havlin}},
  \bibinfo{author}{\bibfnamefont{M.}~\bibnamefont{Schwartz}}, \bibnamefont{and}
  \bibinfo{author}{\bibfnamefont{H.~E.} \bibnamefont{Stanley}},
  \bibinfo{journal}{Phys. Rev. A} \textbf{\bibinfo{volume}{44}},
  \bibinfo{pages}{R2239} (\bibinfo{year}{1991}).

\bibitem[{\citenamefont{Prakash et~al.}(1992)\citenamefont{Prakash, Havlin,
  Schwartz, and Stanley}}]{Prakash1992}
\bibinfo{author}{\bibfnamefont{S.}~\bibnamefont{Prakash}},
  \bibinfo{author}{\bibfnamefont{S.}~\bibnamefont{Havlin}},
  \bibinfo{author}{\bibfnamefont{M.}~\bibnamefont{Schwartz}}, \bibnamefont{and}
  \bibinfo{author}{\bibfnamefont{H.~E.} \bibnamefont{Stanley}},
  \bibinfo{journal}{Phys. Rev. A} \textbf{\bibinfo{volume}{46}},
  \bibinfo{pages}{R1724} (\bibinfo{year}{1992}).

\bibitem[{\citenamefont{Frigo and Johnson}(2005)}]{FFTW05}
\bibinfo{author}{\bibfnamefont{M.}~\bibnamefont{Frigo}} \bibnamefont{and}
  \bibinfo{author}{\bibfnamefont{S.~G.} \bibnamefont{Johnson}},
  \bibinfo{journal}{Proceedings of the IEEE} \textbf{\bibinfo{volume}{93}},
  \bibinfo{pages}{216} (\bibinfo{year}{2005}), \bibinfo{note}{special issue on
  ``Program Generation, Optimization, and Platform Adaptation''}.

\bibitem[{\citenamefont{Bray and Moore}(1985)}]{BrayMoore1985}
\bibinfo{author}{\bibfnamefont{A.~J.} \bibnamefont{Bray}} \bibnamefont{and}
  \bibinfo{author}{\bibfnamefont{M.~A.} \bibnamefont{Moore}},
  \bibinfo{journal}{J. Phys. C: Solid State Phys.}
  \textbf{\bibinfo{volume}{18}}, \bibinfo{pages}{927} (\bibinfo{year}{1985}).

\bibitem[{\citenamefont{Picard and Ratliff}(1975)}]{PicardRatliff1975}
\bibinfo{author}{\bibfnamefont{J.~C.} \bibnamefont{Picard}} \bibnamefont{and}
  \bibinfo{author}{\bibfnamefont{H.~D.} \bibnamefont{Ratliff}},
  \bibinfo{journal}{Networks} \textbf{\bibinfo{volume}{5}},
  \bibinfo{pages}{357} (\bibinfo{year}{1975}).

\bibitem[{\citenamefont{Ogielski}(1986)}]{ogielski1986}
\bibinfo{author}{\bibfnamefont{A.~T.} \bibnamefont{Ogielski}},
  \bibinfo{journal}{Phys. Rev. Lett.} \textbf{\bibinfo{volume}{57}},
  \bibinfo{pages}{1251} (\bibinfo{year}{1986}).

\bibitem[{\citenamefont{Goldberg and Tarjan}(1988)}]{GoldbergTarjan1988}
\bibinfo{author}{\bibfnamefont{A.~V.} \bibnamefont{Goldberg}} \bibnamefont{and}
  \bibinfo{author}{\bibfnamefont{R.~E.} \bibnamefont{Tarjan}},
  \bibinfo{journal}{J. ACM} \textbf{\bibinfo{volume}{35}}, \bibinfo{pages}{921}
  (\bibinfo{year}{1988}), ISSN \bibinfo{issn}{0004-5411}.

\bibitem[{\citenamefont{Hartmann and Rieger}(2001)}]{HartmannRieger2001}
\bibinfo{author}{\bibfnamefont{A.~K.} \bibnamefont{Hartmann}} \bibnamefont{and}
  \bibinfo{author}{\bibfnamefont{H.}~\bibnamefont{Rieger}},
  \emph{\bibinfo{title}{Optimization Algorithms in Physics}}
  (\bibinfo{publisher}{Wiley-VCH, Berlin}, \bibinfo{year}{2001}), ISBN
  \bibinfo{isbn}{978-3-527-40307-3}.

\bibitem[{\citenamefont{Mehlhorn and N\"aher}(1999)}]{leda1999}
\bibinfo{author}{\bibfnamefont{K.}~\bibnamefont{Mehlhorn}} \bibnamefont{and}
  \bibinfo{author}{\bibfnamefont{S.}~\bibnamefont{N\"aher}},
  \emph{\bibinfo{title}{The LEDA Platform of Combinatorial and Geometric
  Computing}} (\bibinfo{publisher}{Cambridge University Press},
  \bibinfo{address}{Cambridge}, \bibinfo{year}{1999}),
  \urlprefix\url{http://www.algorithmic-solutions.de}.

\bibitem[{\citenamefont{Binder}(1981)}]{Binder1981}
\bibinfo{author}{\bibfnamefont{K.}~\bibnamefont{Binder}}, \bibinfo{journal}{Z.
  Phys} \textbf{\bibinfo{volume}{43}}, \bibinfo{pages}{119}
  (\bibinfo{year}{1981}).

\bibitem[{\citenamefont{Melchert}(2009)}]{autoScale2009}
\bibinfo{author}{\bibfnamefont{O.}~\bibnamefont{Melchert}},
  \bibinfo{journal}{Preprint: arXiv:0910.5403v1}  (\bibinfo{year}{2009}).

\bibitem[{\citenamefont{Yeomans}(1993)}]{yeomans1993}
\bibinfo{author}{\bibfnamefont{J.~M.} \bibnamefont{Yeomans}},
  \emph{\bibinfo{title}{Statistical Mechanics of Phase Transitions}}
  (\bibinfo{publisher}{Oxford University Press}, \bibinfo{address}{Oxford},
  \bibinfo{year}{1993}).

\bibitem[{\citenamefont{Efron et~al.}(1993)\citenamefont{Efron, Efron, and
  Tibshirani}}]{bootstrapbook}
\bibinfo{author}{\bibfnamefont{B.}~\bibnamefont{Efron}},
  \bibinfo{author}{\bibfnamefont{B.}~\bibnamefont{Efron}}, \bibnamefont{and}
  \bibinfo{author}{\bibfnamefont{R.~J.} \bibnamefont{Tibshirani}},
  \emph{\bibinfo{title}{An Introduction to the Bootstrap}}
  (\bibinfo{publisher}{Chapman \& HALL/CRC}, \bibinfo{year}{1993}).

\bibitem[{\citenamefont{Hartmann}(2009)}]{Hartmann2009}
\bibinfo{author}{\bibfnamefont{A.~K.} \bibnamefont{Hartmann}},
  \emph{\bibinfo{title}{A Practical Guide To Computer Si\-mu\-la\-tion}}
  (\bibinfo{publisher}{World Scientific Publishing Company},
  \bibinfo{year}{2009}), ISBN \bibinfo{isbn}{978-9812834157}.

\bibitem[{\citenamefont{Rushbrooke}(1963)}]{rushbrooke1963}
\bibinfo{author}{\bibfnamefont{G.~S.} \bibnamefont{Rushbrooke}},
  \bibinfo{journal}{J. of Chem. Phys.} \textbf{\bibinfo{volume}{39}},
  \bibinfo{pages}{842} (\bibinfo{year}{1963}).

\bibitem[{\citenamefont{Wiseman and Domany}(1998)}]{WisemanDomany1998}
\bibinfo{author}{\bibfnamefont{S.}~\bibnamefont{Wiseman}} \bibnamefont{and}
  \bibinfo{author}{\bibfnamefont{E.}~\bibnamefont{Domany}},
  \bibinfo{journal}{Phys. Rev. Lett.} \textbf{\bibinfo{volume}{81}},
  \bibinfo{pages}{22} (\bibinfo{year}{1998}).

\bibitem[{\citenamefont{Rushbrooke}(1965)}]{rushbrooke1965}
\bibinfo{author}{\bibfnamefont{G.~S.} \bibnamefont{Rushbrooke}},
  \bibinfo{journal}{J. of Chem. Phys.} \textbf{\bibinfo{volume}{43}},
  \bibinfo{pages}{3439} (\bibinfo{year}{1965}).

\end{thebibliography}
\end{document}